# Half-metallicity in aluminum-doped zigzag silicene nanoribbons

Yao-Jun Dong[1], Xue-Feng Wang[1,2*], P. Vasilopoulos[3], Ming-Xing Zhai[1], and

Xue-Mei Wu[1,2]

[1]Department of Physics, Soochow University, 1 Shizi Street, Suzhou 215006, China

[2] State Key Laboratory of Functional Materials for Informatics, Shanghai Institute of Microsystem and Information Technology, Chinese Academy of Sciences, 865 Changning Road, Shanghai 200050, China

[3]Concordia University, Department of Physics, 7141 Sherbrooke Ouest, Montreal, QC, Canada, H4B 1R6

\* xf_wang1969@yahoo.com

The spin-dependent electronic structures of aluminum-(Al) doped zigzag silicene nanoribbons (ZSiNRs) are investigated by first-principles calculations. When ZSiNRs are substitutionally doped by a single Al atom on different sites in every three primitive cells, they become half-metallic in some cases, a property that can be used in spintronic devices. More interestingly, spin-down electrons can be transported at the Fermi energy when the Al atom is placed on the sub-edge site. In contrast, spin-up electrons can be transported at the Fermi energy when the ZSiNRs are doped on sites near their center. The magnetic moment on edge is considerably suppressed if the Al atom is doped on edge or near-edge sites. Similar results are obtained for a phosphorus-(P) and boron-(B) doped ZSiNR. When two or more Si atoms are replaced by Al atoms, in general the half-metallic behavior is replaced by a metallic, spin gapless semiconducting or semiconducting one. When a line of six Si atoms, along the ribbon's width, are replaced by Al atoms, the spin resolution of the band structure is suppressed and the system becomes nonmagnetic.

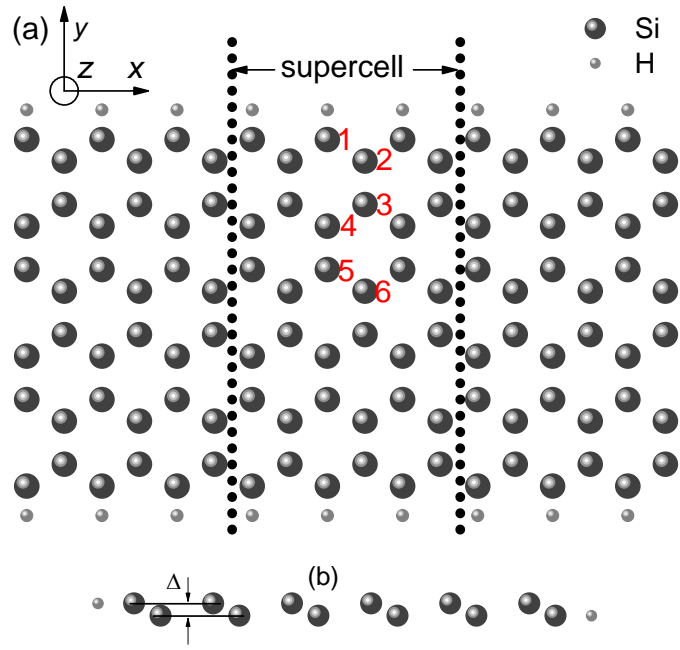

Fig. 1. (a). Supercell of a 6-ZSiNR consisting of three primitive cells along the ribbon (*x*-axis). The numbers, 1, 2, … 6 show the doping sites. The black spheres are the silicon (Si) atoms and the grey ones the hydrogen (H) atoms passivating the dangling Si bonds at the edges. (b) Side view of a 6-ZSiNR; the buckling ∆ is 0.44Å.

**Introduction**

Silicene is a novel two-dimensional (2D) honeycomb structure material formed by silicon (Si) atoms. Compared with the planar graphene, silicene has a low-buckled honeycomb structure [1] and can be easily fabricated [2]. This buckling results in a height difference between the two Si atoms in the primitive cell and leads to a gap that can be controlled by an external electric field since the atoms in a buckled structure are not equivalent. Due to its successful synthesis and its compatibility with silicon-based electronic technology, 2D silicene and silicene nanoribbons SiNRs have potential applications in nanodevices, such as field-effect transistors, and are currently being investigated intensely [1-2].

The electronic structure and transport properties of pristine zigzag SiNRs (ZSiNRs) [3] are similar to those of zigzag graphene nanoribbons (ZGNRs) [4]. There are three main ways to make ZGNRs half-metallic: one is by modifying their edges [5], another is to dope them [6], and a third one is to apply an electric field to them [7]. Since Si materials can be doped by boron (B) or nitrogen (N) atoms, some authors studied the electronic properties of B- or N-doped ZSiNRs and found half-metallic behavior and spin gapless-semiconductor properties [8]. Al-doped bulk Si turns into a *p*-type semiconductor but there is little research on Al-doped SiNRs. The question arises what the electronic properties of Al-doped ZSiNRs are. In order to investigate the influence of Al doping, we replace one Si atom by an Al one on different sites. From the calculation of the band structures, a half-metallic behavior is exhibited for some doping sites. Half-metallicity [9] was first discovered by de Groot *et al.* in 1983 and has been observed in many materials, such as 1D iron-cyclopentadienyl sandwich molecular wires [10], boron nitride nanoribbons [11], silicon nanowires [12], ZGNRs [13] with chemically modified edges, etc. Because of the coexistence of a metallic behavior of electrons for one spin direction with an insulating one for the other spin direction [9], completely spin-polarized half-metallic materials are promising candidates for spintronics applications.

There are three different magnetic configurations for pristine ZSiNRs: antiferromagnetic (AFM), ferromagnetic (FM) and nonmagnetic (NM). Our calculations show that the AFM configuration has the lowest energy in agreement with Ref. [8], that is, it is the ground state of ZSiNRs, while the excited FM state exhibits a metallic behavior [14]. The FM state may be obtained using ferromagnetic electrodes to gain efficient spin injection into ZSiNRs [15]. We use a FM configuration of the atoms between the two edges as the initial one.

In this paper we provide another way of producing half-metallic ZSiNRs by chemically replacing a Si atom by an Al atom on different sites. The doping sites in a 6-ZSiNR are shown Fig. 1(a). When Al is placed on sites 2, the half-metallicity is revealed in spin-down electrons, whereas it is revealed in spin-up electrons when the doping sites are 4 and 5. In Sec. 2 we present the computational model and method and in Sec. 3 the results. A summary follows in Sec. 4.

## 2. Computational model and method

In pristine ZSiNRs the bukling is 0.44 Å and the Si-Si and Si-H bond lengths are about 2.25 Å and 1.5 Å, respectively [1]. In order to investigate the effects of an Al atom replacing periodically a Si one at different sites on the structural and electronic properties of $n$-ZSiNRs, with even width number $n$=6 [16], we consider a system with supercell composed of three primitive unit cells, as shown by the dotted boundaries in Fig. 1. Since 6-ZSiNRs are rotationally symmetric about the $x$ axis, there are only six different Al doping sites. In the simulation process we first perform a geometric optimization for each system using the Atomistix ToolKit (ATK) to adjust the structure parameters until the force felt by each atom becomes less than 0.02 eV/Å. Secondly, the band structure is calculated by the density functional theory (DFT) as implemented in the ATK package [17]. For the exchange-correlation potential we use the generalized gradient approximation with the revised Perdew-Burke-Ernzerhof parametrization. In all calculations we use a basis set of double zeta orbitals plus one polarization orbital to obtain an accurate description of the band structures. A 15Å thick vacuum layer is adopted to separate the ribbons in neighboring supercells, along the $y$ and $z$ axes, by a distance long enough to suppress any coupling between them. In the calculation the energy cutoff is 150 Ry, the mesh grid in $k$ space $1\times1\times100$. To help investigate the half-metallic behavior, we calculated the band structures and the atomic magnetic moments for all doping sites.

## 3. Results and discussion

**A.** *One-atom substitution.* We first present results when an Al atom replaces a Si atom on different sites, labeled 1-6 in Fig. 1.

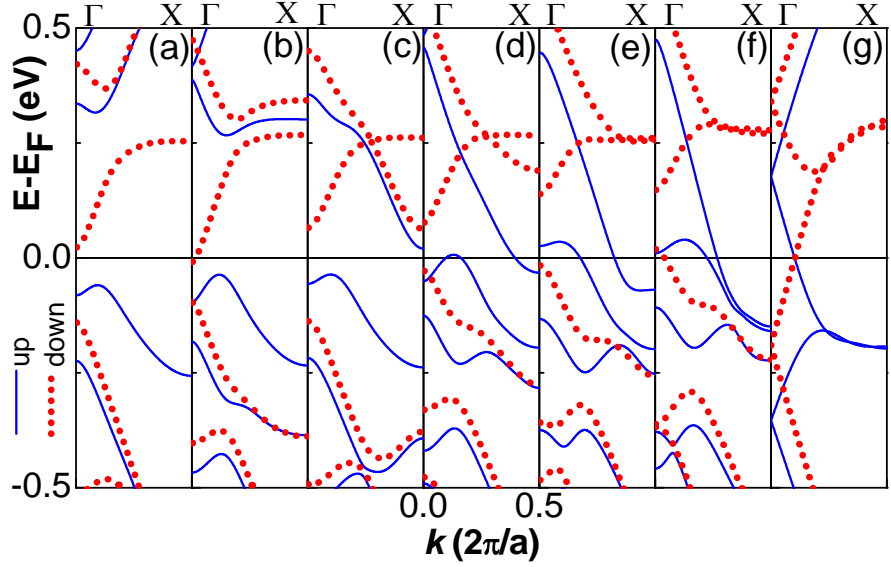

Fig. 2. (Color online). Band structures of a 6-ZSiNR with an Al atom placed on sites 1-6 in panels (a)-(f), respectively, and (g) of a pristine 6-ZSiNR. The blue solid curves are for spin-up electrons and the red dotted ones for spin-down electrons. All energies are measured from the Fermi energy.

The spin-polarized band structures of all Al-doped 6-ZSiNRs, in the FM configuration, shown in Fig. 2, are helpful in understanding and analyzing their half-metallic behavior. As can be seen, the band structure depends considerably on the doping site showing ample variety. In the case of doping site 1, the 6-ZSiNR becomes a semiconductor with spin-down conduction band (CB) bottom and spin-up valence band (VB) top as indicated in Fig. 2(a). As shown in Fig. 2(b), only the CB ($\pi^*$ band) of the spin-down channel crosses the Fermi level while the spin-up band does not in case that the Al atom is placed on site 2. This indicates that for doping site 2 the ribbon becomes half-metallic. When the Al atom is placed on site 3, the material becomes a semiconductor with very narrow indirect spin-up energy gap and direct spin-down energy gap, different from the case of site-1 doping where direct gaps are observed for both spins. If the Al is placed on sites 4 or 5, the doped ribbons are half-metallic as seen in Fig. 2(d) and (e), respectively. Unlike the case of doping on site 2, two CBs of spin-up electrons cross the Fermi level, but the spin-down bands have a gap, which may be very useful in designing spintronic devices. When a Si atom is replaced by Al on site 6, the doped ribbon becomes a normal metal as shown in Fig. 2(f). Compared with the pristine 6-ZSiNR shown in panel (g), when the Al doping site approaches the ribbon's center along the y axis, the band structure is similar to that of the pristine ribbon. In contrast, when the doping site approaches the edge, the π* band is away from the Fermi energy, see Fig. 2(a), (b).

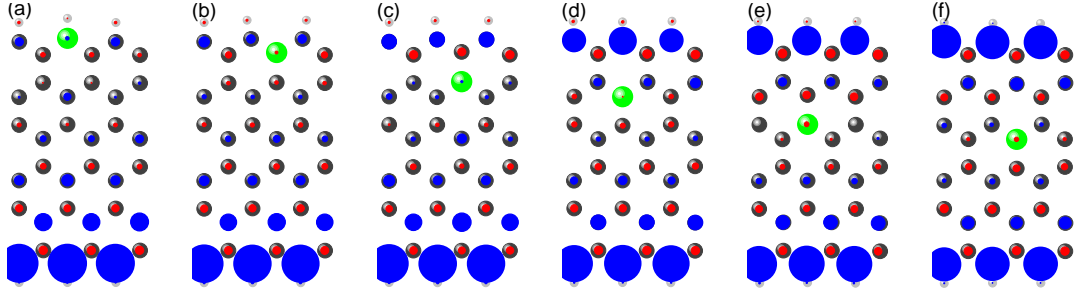

Fig. 3. (Color online). The distribution of atomic magnetic moments (up, down) of an Al-doped 6-ZSiNR: panels (a), (b), (c), (d), (e), and (f) are for sites 1, 2, 3, 4, 5, and 6, respectively. The large green atoms are the Al impurities, the small grey ones the hydrogen atoms, and the black ones the Si atoms. The blue circles show that the net atomic magnetic moment, the difference between numbers of spins up and spins down, is positive, while the red circles show it's negative.

In order to understand the changes in the band structures of the doped 6-ZSiNRs, we show the net atomic magnetic moment of an Al-doped 6-ZSiNR in Fig. 3. Different from the FM configuration of pristine 6-ZSiNRs, the π* bands for the edge states are shifted from the Fermi energy to much higher energies by the Al atom if doped on sites 1, 2, and 3 as indicated in Fig. 2 (a), (b), and (c). Notice that site 1 is on the edge and sites 2 and 3 near the edge of the ribbon. The substituted Al atom acts as a donor atom and injects electrons into the surrounding Si atoms. From the atomic magnetic density point of view, the symmetric spin distribution is broken, especially when an Al atom is placed on sites 1, 2, and 3. The spin polarization is mostly localized on the edges without doping, very similar to that shown in Fig. 3 (f), while it is suppressed when the impurity is on the edge. This is due to the removal of the otherwise partly occupied π* edge states near the Fermi energy. With the doping site moving inside along the *y* axis, the spin polarization reappears as shown in Fig. 3(d), (e), and (f), as the effect of the impurity Al atom on the π* edge states becomes weak and the energies of the states recover to their values in pristine nanoribbons. Our results clearly show that the edge magnetism in ZSiNRs would be greatly suppressed by the substitution of Al atoms on edge sites.

One may wonder whether the results presented so far are specific to an Al dopant. To test that we first performed calculations for a P dopant. The results corresponding to the first six panels (a)-(f) of Fig. 2 are presented in Fig. 4 and the curves are marked as in Fig. 2. As shown, in all cases we observe qualitatively the same half-metallic behavior. The main difference comes from the opposite doping types of Al and P elements. When the P atom is placed on site 2 or 4, only the spin-up valence band (π band) crosses the Fermi energy level and when it is placed on site 5, two spin-down bands cross the Fermi energy as shown in Fig. 4.

Secondly we calculate the electronic properites of 6-ZSiNRs doped by an atom of another group III isoelectronic element B as shown in Fig. 5. Note that some of the results for B doped ZSiNRs have been addressed by some other groups [18]. The doped ribbons are half-metallic when one B atom is doped on sites 1, 2 and 3 as seen in Fig. 5(a), (b) and (c), respectively. Only one spin-down energy band crosses the Fermi level. While being substitutionally doped by B atoms on sites 4, 5, and 6, the ribbons are all metallic as illustrated in Fig. 5(d), (e) and (f), respectively. These results distinguish clearly from those for Al doped nanoribbons presented in Fig. 2.

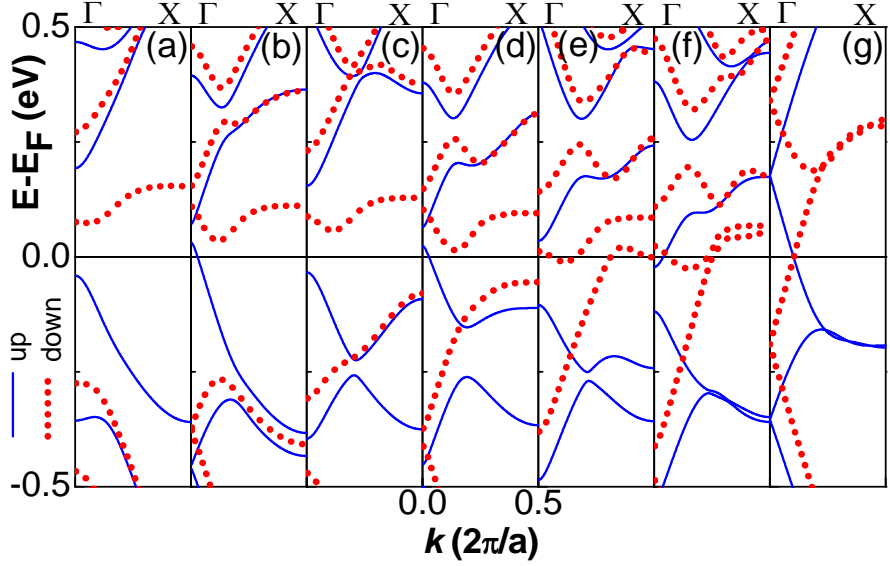

Fig. 4. (Color online). As in Fig. 2 but for a P atom placed on sites 1-6 in panels (a)-(f), respectively. Panel (g) is for the pristine 6-ZSiNR.

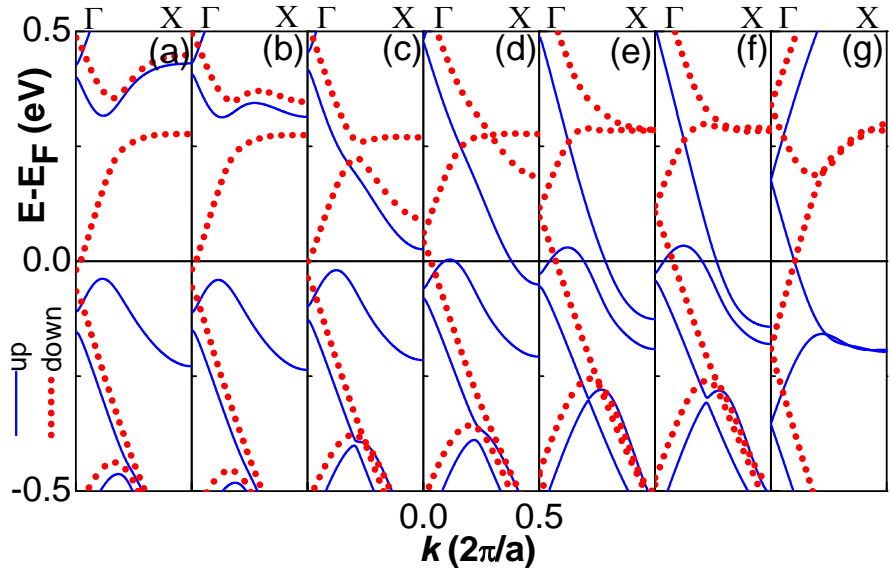

Fig. 5. (Color online). As in Fig. 2 but for a B atom placed on sites 1-6 in panels (a)-(f), respectively. Panel (g) is for the pristine 6-ZSiNR.

*B. Multiple-atom substitution.* We now present results when two or more Al atoms replace Si atoms. First, we show in Fig. 6 the band-structure results when two Al atoms (in green) are placed on sites (a) 1, 2, (b) 1, 4, and (c) 2, 3, respectively. The curves are marked as in Fig. 2. As can be seen, in all cases (a), (b), and (c) we observe a metallic behavior.

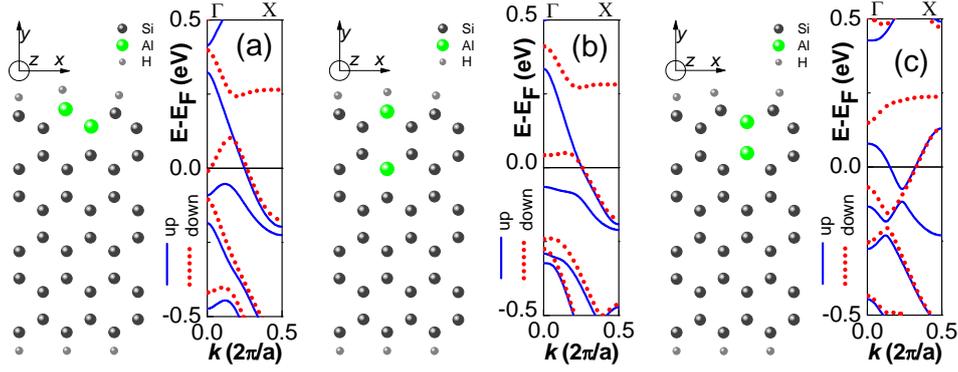

Fig. 6. (Color online). Band structures of a 6-ZSiNR with two Al atoms (in green) placed on sites (a) 1, 2, (b) 1, 4, and (c) 2, 3. The blue solid curves are for spin-up electrons and the red dotted ones for spin-down electrons. All energies are measured from the Fermi energy. The ribbon shows a metallic behavior in all cases (a), (b), and (c).

We continue with the case of three Al atoms replacing three Si atoms on sites (a) 1, 4, 5, and (b) 2, 3, 6. The results are shown in Fig. 7 and the curves are marked as in Fig. 2. Notice that the spin-down energy gap in Fig. 7(a) is very narrow and the result can be considered as a spin gapless semiconductor [19]. In contrast, the energy gaps of both spins are very narrow and comparable in Fig. 7(b) showing narrow semiconductor behavior. When the lines of the three Si atoms are parallel to the ribbon's edges, the results are shown in Fig. 8. As shown, in both cases the 6-ZSiNR behaves as a metal. Notice, however, that the band widths are significantly larger than those in Fig. 7.

Another case of interest is shown in Fig. 9 in which a line of six Al atoms replaces one of six Si atoms. In this case the system remains the mirror symmetry along the central line of the ribbon as in the pristine case and bands are not spin resolved. The behavior of the ribbon changes from a metalic one in (a) to a semiconducting one in (b) when the line in (a) is shifted to the right, as shown in the two supercells.

Finally, we consider the case of a 6-ZSiNR when a ring of six Al atoms replaces a ring of six Si atoms as shown at the center of the supercell. The results are shown in

Fig. 10 and the curves are marked as in Fig. 2. In this case the 6-ZSiNR behaves like a metal.

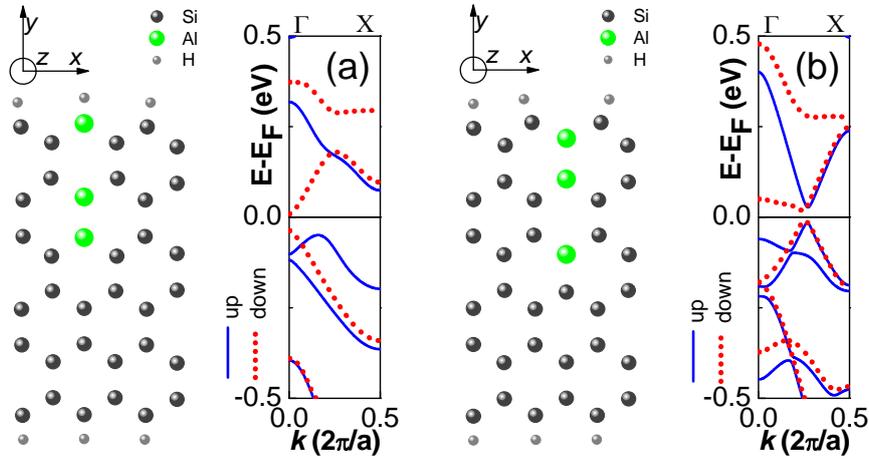

Fig. 7. (Color online). Band structures of a 6-ZSiNR with three Al atoms (in green) repla- cing Si atoms on sites (a) 1, 4, 5, and (b) 2, 3, 6, see the supercells. The blue solid curves are for spin-up electrons and the red dotted ones for spin-down electrons. Notice that the half-metallic behavior in (a) is replaced by a semiconducting one in (b). All energies are measured from the Fermi energy.

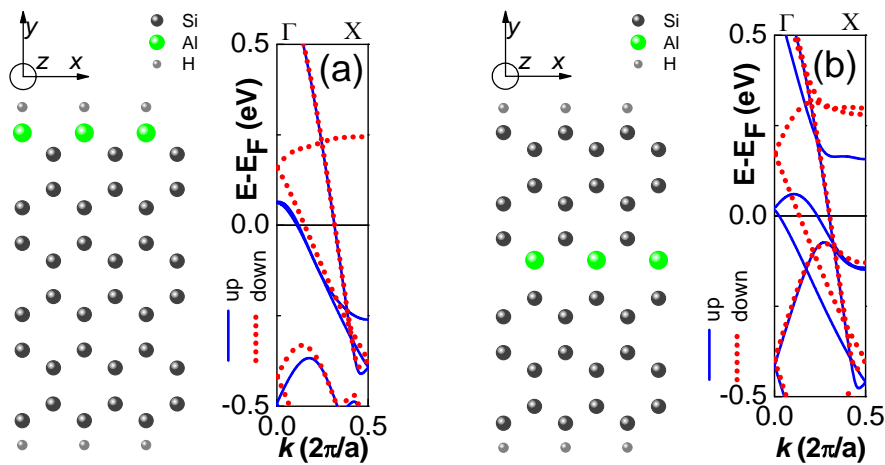

Fig. 8. (Color online). As in Fig. 7 but with three Al atoms (in green) replacing three Si atoms on the horizontal lines, parallel to the edges, that pass through (a) site 1 and (b) site 6, see the supercells. In both cases the

6-ZSiNR behaves as a metal.

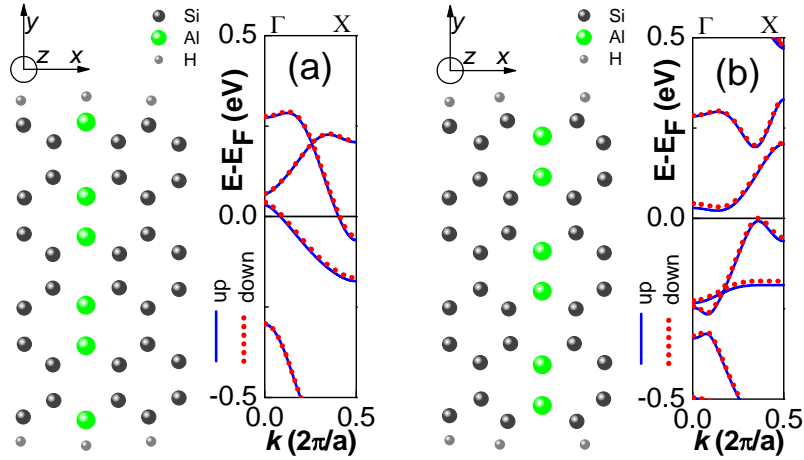

Fig. 9. (Color online). Band structures of a 6-ZSiNR with a line of six Al atoms (in green) re- placing a line of Si atoms as shown on the supercell structures. In either panel there is no difference between spin-up and spin-down electrons. The results in (a) show a metallic behavior, the ones in (b) a semiconducting one. All energies are measured from the Fermi energy.

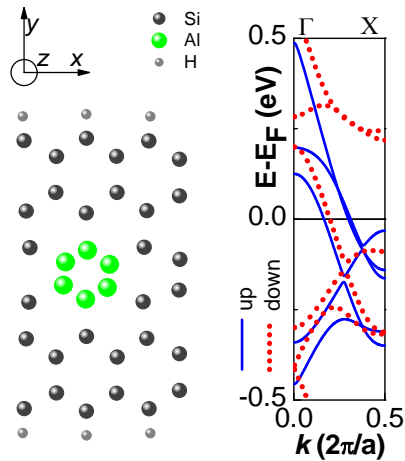

Fig. 10. (Color online). Band structure of a 6-ZSiNR with a ring of six Al atoms (in green) replacing a ring of six Si atoms as shown on the supercell structure. The blue solid curves are for spin-up electrons and the red dotted ones for spin-down electrons. All energies are measured from the Fermi

energy. Notice the metallic behavior of this 6-ZSiNR.

## 4. Summary

Using first-principles calculations, we investigated the band-structure properties of 6-ZSiNRs when one or more Al atoms replace one or more Si atoms on different sites. When an Al atom replaces a single Si one, on sites 2, 4, or 5 of a supercell of three primitive cells, the 6-ZSiNRs show a half-metallic behavior and a 100% spin polarization. This indicates that 6-ZSiNRs, doped by Al on different sites, can provide various designs of spintronic devices. From the evaluated distribution of the atomic magnetic moments of Al-doped 6-ZSiNRs, we infer that the spin polarization is suppressed for certain Al doping sites. For two or more Si atoms replaced, in general a metallic, or semiconducting behavior is observed but depending on the positions of the replaced atoms a gapless spin semiconductor behavior is possible, see Fig. 7(a). Another noteworthy feature is that when a line of six Al atoms replaces one of Si atoms, along the ribbon's width, the spin resolution of the band structure is suppressed due to the geometry symmetry, see Fig. 9. Similar results are obtained for P-doped 6-ZSiNRs, cf. Fig. 4, but with the spin orientation reversed. The 6-ZSiNRs appear half-metallic when being doped by element B on site 1, 2, or 3.

## Acknowledgments

This work was supported by the National Natural Science Foundation in China (Grant Nos. 11074182 and 91121021), a Program for graduates Research & Innovation in University of Jiangsu Province (CXZZ13_0797) and by the Canadian NSERC Grant No. OGP0121756. It is also partially supported by the Qing Lan Project funded by the Priority Academic Program Development of Jiangsu Higher Education Institutions.